\documentclass[aps,prl,twocolumn,a4paper,superscriptaddress,amsmath,amssymb]{revtex4}
\usepackage[dvips]{graphicx}
\usepackage[dvips]{color}

\renewcommand{\vec}[1]{\mathbf #1} \renewcommand{\i}{\mathrm i}
\renewcommand{\exp}{\mathrm e} \newcommand{\erf}{\mathrm {erf}}

\renewcommand{\d}{\mathrm d} \renewcommand{\r}{\vec r}
\renewcommand{\Im}[1]{\text{Im}\left\{ #1\right\}}
\renewcommand{\Re}[1]{\text{Re}\left\{ #1\right\}}

\begin{document}

\title{Spin transport in diffusive superconductors}

\author{Jan Petter Morten}
\email{jan.morten@phys.ntnu.no}
\author{Arne Brataas}
\affiliation{Department of Physics, Norwegian University of Science
and Technology, 7491 Trondheim, Norway}

\author{Wolfgang Belzig}
\affiliation{Department of Physics and Astronomy, University of Basel,
Klingelbergstrasse 82, 4056 Basel, Switzerland}

\date{Received 28 January 2004}

\begin{abstract}
We employ the Keldysh formalism in the quasiclassical approximation to
study transport in a diffusive superconductor. The resulting $4\times
4$ transport equations describe the flow of charge and energy as well
as the corresponding flow of spin and spin energy. Spin-flip
scattering due to magnetic impurities is included. We find that the
spin-flip length is renormalized in the superconducting case and
propose an experimental system to measure the spin-accumulation in a
superconductor.
\end{abstract}


\maketitle

Manipulation of spin-polarized currents can be used to study
fundamental transport processes and might also provide new
functionality in electronic devices. In ferromagnets (F), the current
is spin-polarized due to the spin-dependent density of states and the
spin-dependent scattering potentials. In contrast, in $s$-wave
superconductors (S), electrons with spin up and spin down and opposite
momentum form Cooper pairs with no net spin.  Nano-scale
superconductors therefore display strikingly different properties when
driven out of equilibrium by spin transport than by charge transport.

Most of the recent activities on the transport properties of F/S
junctions have studied effects caused by the physical properties on
the F side of the junction. The zero spin Cooper pairs prevent
spin-polarized electrons to flow into S. Consequently, a spin
polarized current from F injected into S can result in nonequilibrium
spin accumulation near the F/S interface. The competition between
electron-hole correlations and spin accumulation on the F side has
recently attracted considerable interest\cite{fs}.  Possible influence
of the ferromagnetic order parameter on the superconductor has
received less attention. Singlet pairing does not allow a spin
accumulation in the superconductor. Consequently, spin accumulation
can reduce the superconducting gap and change the transport properties
both for transport via quasiparticles and for the supercurrent.
Experimentally, spin transport in diffusive S has recently been
studied \cite{Gu:prb02}. Here, the reduced quasiparticle penetration
due to spin accumulation results in loss of spin memory which can be
measured as a decreased magnetoresistance.

Although the theory of nonequilibrium superconductivity is widely used
and developed, it has to the best of our knowledge not been completely
generalized to study spin transport. In this work we thus use the
Keldysh formalism and the quasiclassical approximation
\cite{Belzig:sm99,Kopnin:01,Rammer:rmp86} to rigorously obtain a set
of equations describing the transport of charge and energy in a
diffusive weak coupling S, as well as the transport of spin.  This
will describe the penetration of spins into S and the associated
suppression of the superconducting order parameter. Our description of
the transport properties will be based on a 4$\times$4 matrix equation
formalism to include spin accumulation as well as electron-hole
correlations. Spin-flip scattering from magnetic impurities is
included as the dominant spin relaxation process inside the
superconductor. We find that the spin-flip length is renormalized in
the BCS case, and propose an experimental system to measure the
properties resulting from the superconducting correlations.  Many, but
not all, experimental systems involving spin-transport in
superconductors are in the {\it elastic} transport regime
\cite{Tserkovnyak:prb02}, which is considered here. Complementary
studies based on the Boltzmann equation for spin-transport by
quasiparticles in the {\it inelastic} transport regime have recently
been published \cite{Maekawa}. Note that spin-injection is
qualitatively different in these opposite transport regimes due to the
strong energy-dependence of quasiparticle flow in superconductors
\cite{Maekawa}.

Let us now outline the derivation of our main results. We use natural
units so that $\hbar=k_\text{B}=1$, and the electron charge is
$e=-|e|$. To describe the out-of-equilibrium electron-hole
correlations as well as spin accumulation, we define the Keldysh
Green's function as
\begin{equation}
  \hat{G}^\text{K}_{ij}(1,2)
  =\mathop{\sum}_k(-\i)\left(\hat{\rho}_3\right)_{ik}\left<[\left(\psi(1)\right)_k,\left(\psi^\dag(2)\right)_j]_{-}\right>,
\end{equation}
where
$\psi=[\psi_\uparrow,\psi_\downarrow,\psi_\uparrow^\dag,\psi_\downarrow^\dag]^\text{T}$
is a four-vector and $\psi^\dag$ the corresponding adjoint vector. The
matrix $\hat{\rho}_3$ is the third Pauli matrix generalized to
4$\times$4 space, $\hat{\rho}_3={\rm diag}(1,1,-1,-1)$. The
coordinates are $1=(\r_1,t_1)$ and $2=(\r_2,t_2)$. Similarly, we
define $4\times 4$ retarded and advanced Green's functions
($\hat{G}^\text{R},~\hat{G}^\text{A}$) in spin- and particle-hole
space. $4 \times 4$ matrices are denoted by a ``hat'' superscript. A
compact notation can be obtained by construction of an 8$\times$8
matrix in the Keldysh space (denoted by a ``check'' superscript)
\cite{Rammer:rmp86}.

The quasiclassical Green's function is defined by
$\check{g}(\vec{R},T,\vec{p}_\text{F},E)=\frac{\i}{\pi}\int\d\xi_{\vec{p}}\check{G}(\vec{R},T,\vec{p},E)$.
This function is determined by the Eilenberger equation which in the
mixed representation for a stationary state can be written
\begin{equation}
  \left[E\hat{\rho}_3+\i\frac{\vec{p}}{m}\cdot\hat{\boldsymbol{\partial}}-e\phi\hat{1}-\hat{\Delta}-\check{\sigma},\,\check{g}\right]_-=0.
\end{equation}
Here $\boldsymbol{\hat{\partial}}=\nabla\hat{1}-\i
e\vec{A}\hat{\rho}_3$ is the gauge invariant derivative, $\hat{1}$ is
the 4$\times$4 unit matrix, $\phi$ is the electromagnetic scalar
potential, $\hat{\Delta}$ contains the superconducting gap and
$\check{\sigma}$ is the self-energy due to elastic impurity scattering
and spin-flip scattering by magnetic impurities in quasiclassical
approximation. In the case of strong impurity scattering (dirty limit)
transport is diffusive. Expansion of the quasiclassical Green's
function in spherical harmonics then gives the Usadel equations. The
symmetries and normalization of the Green's function allows for a
parameterization of the quasiclassical, retarded component
\cite{Belzig:sm99}
\begin{equation}
  \hat{g}_s^\text{R}=\begin{pmatrix}\bar{1}\cosh(\theta) &
  \i\bar{\tau}_2\sinh(\theta)\exp^{\i\chi} \\
  \i\bar{\tau}_2\sinh(\theta)\exp^{-\i\chi} & -\bar{1}\cosh(\theta)
  \end{pmatrix},
\end{equation}
where $\bar{1}$ is the 2$\times$2 unit matrix, $\bar{\tau}_2$ is the
second Pauli matrix and $\theta$ and $\chi$ are position and energy
dependent functions. We assume colinear magnetizations along the
$z$-axis and $s$-wave singlet superconducting state. We choose a gauge
where the superconducting order parameter $\Delta$ is real and
positive, and then the supercurrent is contained in the
electromagnetic vector potential $\vec{A}$ and the chemical potential
of the Cooper pairs is included in $\phi$. Inspection of the
self-consistency relation for $\Delta$ reveals that $\chi=0,\pi$
depending on the boundary conditions. This ansatz simplifies the
calculations considerably. The advanced Green's function is related to
the retarded through
$\hat{g}^\text{A}=-\left[\hat{\rho}_3\hat{g}^\text{R}\hat{\rho}_3\right]^\dag$.
Because of normalization, the Keldysh Green's function can be
expressed as
$\hat{g}^\text{K}=\hat{g}^\text{R}\hat{h}-\hat{h}\hat{g}^\text{A}$
where $\hat{h}$ is a diagonal distribution function matrix.
\par
We will now consider a stationary state. A kinetic equation can be
derived from the Usadel equations if we include Keldysh components.
The important quantities are the physical particle and energy currents
(including particles and holes), which we will denote by
$\vec{j}_\text{T}$ and $\vec{j}_\text{L}$ respectively, with the
corresponding distribution functions carrying the same indices,
$h_\text{T}$ and $h_\text{L}$ \cite{Belzig:sm99}. The physical spin
current is denoted $\vec{j}_\text{TS}$ and the spin energy current
$\vec{j}_\text{LS}$, with distribution functions $h_\text{TS}$ and
$h_\text{LS}$. The spin-resolved distribution functions can be
expressed by the particle distribution function as
$h_\text{TS(LS)}=-(f_\uparrow(E)-f_\downarrow(E))/2-(+)(f_\uparrow(-E)-f_\downarrow(-E))/2$.
The current components $\vec{j}_\text{T}$ {\it etc.} are spectral
quantities, and the total charge current is given as an integral
$\vec{j}_\text{charge}(\vec{r},t)=|e|N_0\int_{-\infty}^{\infty}\d
E\vec{j}_\text{T}(\vec{r},t,E)$, and the spin current is obtained by a
similar integral of $\vec{j}_\text{TS}$. Energy current is given by
$\vec{j}_\text{energy}(\vec{r},t)=|e|N_0\int_{-\infty}^{\infty}\d
E\,E\vec{j}_\text{L}(\vec{r},t,E)$, and the difference in energy
current carried by opposite spins by a similar integral of
$\vec{j}_\text{LS}$.

 The
equilibrium solutions for the distribution functions are
$h_\text{L,0}=\tanh\left(\beta E/2\right)$ and
$h_\text{T,0}=h_\text{LS,0}=h_\text{TS,0}=0$. We derive kinetic
equations and find,
\begin{subequations}
  \label{eq:kinetic}
  \begin{align}
    \nabla\cdot\vec{j}_\text{L}&=0,\\
    \nabla\cdot\vec{j}_\text{T}&=-2|\Delta|\alpha_\text{TT}\,h_\text{T},\\
    \nabla\cdot\vec{j}_\text{LS}&=-\left(2|\Delta|\alpha_\text{TT}+\frac{1}{\tau_\text{sf}}\alpha_\text{LSLS}\right)h_\text{LS},\label{eq:LSkinetic}\\
    \nabla\cdot\vec{j}_\text{TS}&=-\frac{1}{\tau_\text{sf}}\alpha_\text{TSTS}\,h_\text{TS}.\label{eq:TSkinetic}
  \end{align}
\end{subequations}
The right-hand side terms represent renormalized scattering because of superconductivity:
\begin{subequations}
  \label{eq:alpha}
  \begin{align}
    \alpha_\text{TT}=&\Im{\sinh(\theta)},\\
    \alpha_\text{LSLS}=&\left(\Re{\cosh(\theta)}\right)^2-\left(\Im{\sinh(\theta)}\right)^2,\\
    \alpha_\text{TSTS}=&\left(\Re{\cosh(\theta)}\right)^2+\left(\Re{\sinh(\theta)}\right)^2.
  \end{align}
\end{subequations}
The $|\Delta|\alpha_\text{TT}$ terms on the right hand side in
\eqref{eq:kinetic} are due to conversion of quasiparticle current into
supercurrent, and the
$\alpha_\text{LSLS}/\tau_\text{sf},~\alpha_\text{TSTS}/\tau_\text{sf}$
terms are due to spin-flips. The spin-flip time in the normal state is
$\tau_\text{sf}^{-1}=8\pi n_\text{sf}N_0 S(S+1)|v_\text{sf}|^2/3$,
where $n_\text{sf}$ is the magnetic impurity density, $N_0$ the
density of states at the Fermi level, $S$ the impurity spin quantum
number and $v_\text{sf}$ is the Fourier transformed spin-flip impurity
potential. We assume isotropic scattering. Our definition of
$\tau_\text{sf}$ differs from the usual spin-flip lifetime by a
renormalization factor $4/3$. This definition reproduces the diffusion
equation with a spin-flip length
$l_\text{sf}^\text{(N)}=\sqrt{D\tau_\text{sf}}$ in the normal state.
Thus there is a difference between the spin-flip lifetime measured in
{\it e.g.} electron spin resonance and spin-flip transport time.

We introduce generalized energy-dependent diffusion coefficients
\begin{subequations}
\begin{align}
  D_\text{L}=&D\left[\left(\Re{\cosh(\theta)}\right)^2-\left(\Re{\sinh(\theta)}\right)^2\right],\\
  D_\text{T}=&D\left[\left(\Re{\cosh(\theta)}\right)^2+\left(\Im{\sinh(\theta)}\right)^2\right],
\end{align}
\end{subequations}
where $D=\tau v_\text{F}^2/3$ is the diffusion constant. The currents
can then be expressed as
\begin{subequations}
  \label{eq:juttrykk}
  \begin{align}
    \vec{j}_\text{L}=&-D_\text{L}\nabla h_\text{L}+\Im{\vec{j}_\text{E}}h_\text{T},\label{eq:Lcurrent}\\
    \vec{j}_\text{T}=&-D_\text{T}\nabla h_\text{T}+\Im{\vec{j}_\text{E}}h_\text{L},\\
    \vec{j}_\text{LS}=&-D_\text{T}\nabla h_\text{LS}+\Im{\vec{j}_\text{E}}h_\text{TS},\\
    \vec{j}_\text{TS}=&-D_\text{L}\nabla h_\text{TS}+\Im{\vec{j}_\text{E}}h_\text{LS}.\label{eq:TScurrent}
  \end{align}
\end{subequations}
Here we have defined the spectral supercurrent as
$\vec{j}_E=D(\nabla\chi-2e\vec{A})\sinh^2(\theta)$. The
self-consistency relation is
\begin{align}
  \Delta(\r)=&-\frac{1}{2}\,\mathrm{sgn}(\Delta_0)N_0\lambda\int_{-\infty}^{\infty}\d E\sinh\left(\theta\right)h_\text{L},
\end{align}
where the factor $\mathrm{sgn}(\Delta_0)$ is determined from the
boundary condition to give the correct sign and $\lambda$ is the
interaction parameter. The complex part of this equation is neglected
as a consequence of charge conservation \cite{Schmid:nato81}.
\par
The functions $\theta$ and $\chi$ are determined by the retarded
components of the Usadel equation. We obtain
\begin{align}
  \label{eq:usadel1}
  \nabla\cdot\vec{j}_E=&0,\\
  \label{eq:usadel2}
  D\left[\nabla^2\theta-\frac{1}{2}(\nabla\chi-2e\vec{A})^2\sinh(2\theta)\right]=&-2\i E\sinh(\theta)\nonumber\\-2\i\cosh(\theta)|\Delta|+\frac{3}{4}&\frac{1}{\tau_\text{sf}}\sinh(2\theta),
\end{align}
where Eq. \eqref{eq:usadel1} implies that the spectral supercurrent is
conserved. In addition we have the following symmetry conditions,
$\theta^*(-E)=-\theta(E), ~\chi^*(-E)=\chi(E)$. Equations
\eqref{eq:kinetic}-\eqref{eq:usadel2} determine all transport
properties of S.

In general, in a hybrid F/S system, the superconductor cannot be
described as in terms of BCS-formulas close to the F/S interface due
to the proximity effect. Nevertheless, to gain insight into the
physics implied by the above-mentioned formulas let us now consider
the limit of a homogeneous BCS superconductor, and select $\chi=0$.
This is relevant for the proposed experiment below. For energies
$|E|<|\Delta|$ $\alpha_\text{TT}=\Delta/ \sqrt{\Delta^2-E^2}$ and the
spin-flip renormalization factors are
$\alpha_\text{TSTS}=0,~\alpha_\text{LSLS}=-\Delta^2/(\Delta^2-E^2)$.
The generalized diffusion constant $D_\text{L}=0$ while
$D_\text{T}=D\Delta^2/(\Delta^2-E^2)$. From Eq. \eqref{eq:Lcurrent}
this means that there is no energy current carried by quasiparticles
with energy $|E|<|\Delta|$. Gap scattering for quasiparticle energies
below the superconducting gap corresponds to a transformation of the
charge current ($\vec{j}_\text{T}$) into supercurrent. Such scattering
is not possible for the physical spin current ($\vec{j}_\text{TS}$).
Consequently, in the absence of spin-flip scattering the quasiparticle
spin-current into the superconductor vanishes for $|E|<|\Delta|$ since
$D_\text{L}=0$ in Eq. \eqref{eq:TScurrent} and $\alpha_\text{TSTS}=0$
in the kinetic equation \eqref{eq:TSkinetic}. Note that this result
relies on the fact that there are different effective diffusion
coefficients for charge current ($D_\text{T}$) and for spin current
($D_\text{L}$). We also observe that the term $\alpha_\text{LSLS}$ is
{\it negative} below the gap, acting as a source of spin energy.
\par
Above the gap $(|E|>|\Delta|)$ the factor $\alpha_\text{TT}$ vanishes
while
$\alpha_\text{LSLS}=E^2/(E^2-\Delta^2),~\alpha_\text{TSTS}=(E^2+\Delta^2)/(E^2-\Delta^2)$.
For the generalized diffusion coefficients we find that $D_\text{L}=D$
and $D_\text{T}=DE^2/(E^2-\Delta^2)$. Now consider the kinetic
equations in the BCS case. A charge current carried by quasiparticles
with energy $|E|>|\Delta|$ can propagate into S. For quasiparticles at
$|E|>|\Delta|$ we see that there is no renormalization for the
spin-energy diffusion length in Eq. \eqref{eq:LSkinetic}, whereas
the spin diffusion length in Equation \eqref{eq:TSkinetic} has an
energy dependent renormalization factor which diverges for energies
$|E|=|\Delta|$ causing massive spin-flip scattering.

\begin{figure}[hh]
  \begin{picture}(0,0)%
\includegraphics{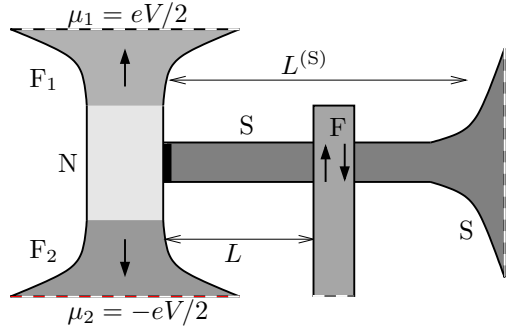}%
\end{picture}%
\setlength{\unitlength}{3158sp}%
\begingroup\makeatletter\ifx\SetFigFont\undefined%
\gdef\SetFigFont#1#2#3#4#5{%
  \reset@font\fontsize{#1}{#2pt}%
  \fontfamily{#3}\fontseries{#4}\fontshape{#5}%
  \selectfont}%
\fi\endgroup%
\begin{picture}(3924,2638)(1779,-9983)
\put(1951,-9361){\makebox(0,0)[lb]{\smash{\SetFigFont{10}{12.0}{\familydefault}{\mddefault}{\updefault}{\color[rgb]{0,0,0}F$_2$}%
}}}
\put(1951,-8011){\makebox(0,0)[lb]{\smash{\SetFigFont{10}{12.0}{\familydefault}{\mddefault}{\updefault}{\color[rgb]{0,0,0}F$_1$}%
}}}
\put(2176,-8686){\makebox(0,0)[lb]{\smash{\SetFigFont{10}{12.0}{\familydefault}{\mddefault}{\updefault}{\color[rgb]{0,0,0}N}%
}}}
\put(2251,-7501){\makebox(0,0)[lb]{\smash{\SetFigFont{10}{12.0}{\familydefault}{\mddefault}{\updefault}{\color[rgb]{0,0,0}$\mu_1=eV/2$}%
}}}
\put(2251,-9841){\makebox(0,0)[lb]{\smash{\SetFigFont{10}{12.0}{\familydefault}{\mddefault}{\updefault}{\color[rgb]{0,0,0}$\mu_2=-eV/2$}%
}}}
\put(3601,-8386){\makebox(0,0)[lb]{\smash{\SetFigFont{10}{12.0}{\familydefault}{\mddefault}{\updefault}{\color[rgb]{0,0,0}S}%
}}}
\put(5326,-9211){\makebox(0,0)[lb]{\smash{\SetFigFont{10}{12.0}{\familydefault}{\mddefault}{\updefault}{\color[rgb]{0,0,0}S}%
}}}
\put(4306,-8396){\makebox(0,0)[lb]{\smash{\SetFigFont{10}{12.0}{\familydefault}{\mddefault}{\updefault}{\color[rgb]{0,0,0}F}%
}}}
\put(3481,-9391){\makebox(0,0)[lb]{\smash{\SetFigFont{10}{12.0}{\familydefault}{\mddefault}{\updefault}{\color[rgb]{0,0,0}$L$}%
}}}
\put(3926,-7911){\makebox(0,0)[lb]{\smash{\SetFigFont{10}{12.0}{\familydefault}{\mddefault}{\updefault}{\color[rgb]{0,0,0}$L^{(\text{S})}$}%
}}}
\end{picture}
  \caption{\label{fig:system}Spin battery connected to a superconductor. The thick solid line indicates a tunnel barrier.}
\end{figure}

We will now apply this formalism to study spin diffusion, and
demonstrate the significance of the renormalization of the spin
diffusion length. Experimental studies of spin accumulation and spin
injection has recently been performed \cite{spininjection} in
metallic spin valves. The spin accumulation in the physically
different {\it inelastic} regime for a superconductor in this
experimental system has also been calculated theoretically
\cite{Maekawa}. We will consider the simplified geometry shown
in Figure \ref{fig:system}, where there is no charge transport in the
superconductor, and calculate the spin accumulation signal in the {\it
  elastic} regime.
The F$_1$/N/F$_2$ systems act as a spin-battery which is connected via
a tunnel junction to the superconductor. A voltage bias between F$_1$
and F$_2$ induces a spin accumulation that can flow into S. The
superconducting wire is connected to an S reservoir in equilibrium BCS
state by a good metallic contact at distance $L^{(\text{S})}$ from the
N/S interface. On top of the S wire there is a ferromagnet connected
by tunnel barrier which upon switching of the magnetization direction
acts as a detector for the spin signal. Measurement of the relative
voltage of this electrode between parallel and antiparallel (with
respect to the top F reservoir) magnetization gives
$\Delta\mu=\mu^{(\text{P})}-\mu^{(\text{AP})}$ which describes the
difference between electrochemical potential of spin-up and spin-down
quasiparticles located a distance $L$ from the N/S interface. This
quantity can be calculated $\Delta\mu=-\int_{-\infty}^{\infty}\d E
P^{(\text{D})}h_\text{TS}(L,E)$, where $P^{(\text{D})}$ is the spin
polarization of the tunnel barrier between S and the F detector. We
assume a homogeneous order parameter and BCS spectral properties in
the S wire since there are tunnel barriers between the N,F and S
elements and perturbation from current and spin-flip is weak.
\par
We can express the difference between the spin-up and spin-down
distribution functions in N close to S as $\Delta f^\text{(N)}\equiv
f_\uparrow^{\text{(N)}}-f_\downarrow^{\text{(N)}}=P^{(\text{FN})}(f(E-eV/2)-f(E+eV/2))$,
where
$P^{(\text{FN})}=\left(G_\text{maj}-G_\text{min}\right)/\left(G_\text{maj}+G_\text{min}\right)$
is the spin polarization between the F reservoirs and N, $f(E\pm
eV/2)$ is the Fermi-Dirac distributions in the F reservoirs and
$G_\text{maj(min)}$ is the conductance of majority (minority) spin
electrons from ferromagnetic reservoir to the middle of N. There is
thus no charge current or supercurrent anywhere in S, however there
may be a spin-current. Equation \eqref{eq:TScurrent} states that there
is no spin-current for energies below the gap, thus for these energies
the N/S interface is effectively insulating. Since the S wire is
connected to a reservoir in the other end for $|E|<\Delta$ the spin
distribution function equals the equilibrium value $h_\text{TS}=0$. We
solve the TS kinetic equation \eqref{eq:TSkinetic} for energies
$|E|>\Delta$. This equation reduces to a diffusion equation with
renormalized spin-flip length
$l_\text{sf}^{(\text{S})}(E)=l_\text{sf}\sqrt{(E^2-\Delta^2)/(E^2+\Delta^2)}$,
where $l_\text{sf}=\sqrt{D\tau_\text{sf}}$ is the normal state
spin-flip length. The boundary condition at the S reservoir is that
the distribution function attains the equilibrium value, and at the
S/N interface we match at each energy the tunnel spin current to the
spin current inside S, $|e|N_0\vec{j}_\text{TS}$. We assume that
$L^{(\text{S})}/l_\text{sf}^{(\text{S})}\gg 1$ which is a relevant
physical situation.
\par
The position and energy dependent solution $h_\text{TS}$ is
substituted into the expression for the measured difference in
electrochemical potential for parallel and antiparallel configuration,
and we obtain
\begin{align}
\Delta\mu=2P^{(\text{D})} \int_\Delta^\infty\d E\, \Delta
f^\text{(N)}\,
\exp^{-L/l_\text{sf}^{(\text{S})}}\frac{R_\text{sf}^{(\text{S})}}{R_\text{sf}^{(\text{S})}+R^{(\text{I})}},
\end{align}
where $R^{(\text{I})}(E)=1/(|T|^2N_\text{BCS}(E)N_0)$ is the
resistance of the N/S tunnel barrier, $|T|$ is the tunneling matrix
element, $N_\text{BCS}(E)$ is the BCS density of states,
$R_\text{sf}^{(\text{S})}(E)=l_\text{sf}^{(\text{S})}(E)\rho/A$ is the
resistance of the S wire within a spin-flip length and $\rho$ is the
resistivity of the material in S when in the normal state
($T>T_c$). This result can be understood as follows. The spin accumulation close to the tunnel interface is exponentially attenuated
by spin-flip scattering in S. The spin signal is also decreased by the
tunnel resistance, and since spin current is strongly suppressed for
energies $|E|<\Delta$ only quasiparticles with energies higher than
the gap contribute. The effective total resistance is a series of the
tunnel interface resistance with the resistance of S within one
spin-flip length.

We will now consider some simplified limits for the quantity
$\Delta\mu$ defined above. In the normal state where
$\Delta\rightarrow 0$ we find that
$\Delta\mu/eV=2P^{\text{(D)}}P^{\text{(FN)}}R_\text{sf}^{\text{(S)}}{\rm
exp}(-L/l_\text{sf})/(R_\text{sf}^{\text{(S)}}+R^\text{(I)})$ where
$R_\text{sf}^{\text{(S)}}$ and $R^\text{(I)}$ assume their normal
state (energy independent) values. At $k_\text{B}T\ll\Delta$ the
signal measured by $\Delta\mu$ vanishes when the bias is lower than
the energy gap $eV<\Delta$ since spin current is suppressed for
quasiparticles below the gap. For higher bias, $eV>\Delta$, and at
zero temperature when the bulk resistance dominates,
$R_\text{sf}^{(\text{S})}\gg R^{(\text{I})},$ an approximate solution
is
$\Delta\mu=2P^{\text{(D)}}P^{\text{(FN)}}\Delta\exp^{-L/l_\text{sf}}\{\exp^{-Lr^2/2l_\text{sf}}/r-\exp^{-L/2l_\text{sf}}+\sqrt{\pi
L/2l_\text{sf}}(\erf[r\sqrt{L/2l_\text{sf}}]-\erf[\sqrt{L/2l_\text{sf}}])\},$
where $r=2\Delta/eV$. In this case the relation between the energy gap
and the bias determines the magnitude of the spin signal, and the
exponential decrease of the signal.

\begin{figure}[hh]
  \begin{picture}(0,0)%
    \includegraphics{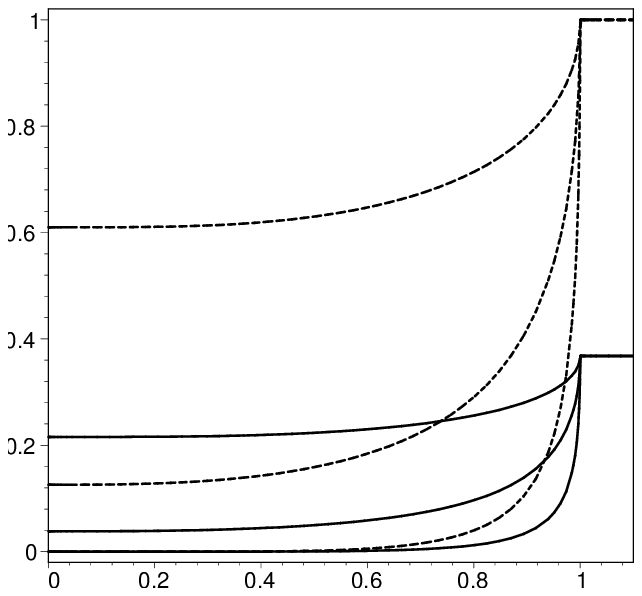}%
  \end{picture}%
  \setlength{\unitlength}{1776sp}%
  \begingroup\makeatletter\ifx\SetFigFont\undefined%
  \gdef\SetFigFont#1#2#3#4#5{%
    \reset@font\fontsize{#1}{#2pt}%
    \fontfamily{#3}\fontseries{#4}\fontshape{#5}%
    \selectfont}%
  \fi\endgroup%
  \begin{picture}(7117,6508)(-449,-5669)
    \put(2851,-5611){\makebox(0,0)[lb]{\smash{\SetFigFont{5}{6.0}{\familydefault}{\mddefault}{\updefault}{\color[rgb]{0,0,0}\normalsize $T/T_\text{c}$}%
        }}}
    \put(-449,-2611){\makebox(0,0)[lb]{\smash{\SetFigFont{5}{6.0}{\familydefault}{\mddefault}{\updefault}{\color[rgb]{0,0,0}\normalsize \rotatebox{90}{$\Delta\mu/eV$}}%
        }}}
  \end{picture}
  \caption{\label{fig:deltamu}Temperature dependence of $\Delta\mu/eV$. We use $R_\text{sf}^{\text{(S)}}=R^\text{(I)}$ in the normal state. For the dotted lines $L/l_\text{sf}=6$, and for the solid lines $L/l_\text{sf}=7$. The bias $eV$ is $0.1\Delta(T=0),~3\Delta(T=0),~10\Delta(T=0)$ for the lower curve to the higher curve, respectively.}
\end{figure}

The temperature dependence of $\Delta \mu$ in the general case is given by
decrease from a constant value above $T_c$ as the temperature
approaches zero. An example of this behavior is shown in Figure
\ref{fig:deltamu}. Here we have used the approximate temperature
dependence $\Delta=1.76 T_c\tanh(1.74\sqrt{T_c/T-1})$.
Our calculations show that the spin signal decreases due to
superconducting correlations. For a large energy gap the spin accumulation vanishes completely at low temperatures. These effects
can be explained by suppressed subgap spin-current and massive
spin-flip at energies close to the gap because of the superconducting
correlations.

In conclusion, we have presented a formalism to describe elastic spin
transport in superconductors with spin-flip scattering. We find
different effective diffusion coefficients for charge- and
spin-current. The spin-flip length is renormalized in the
superconducting case, and at energies close to the gap there is
massive spin-flip. As an illustration we compute
the difference in electrochemical potential due to spin accumulation
in an experiment sensitive to the renormalization of
spin-flip length.

\begin{acknowledgments}
  This work was supported in part by The Research Council of Norway,
  NANOMAT Grants No. 158518/431 and 158547/431, RTN Spintronics, the
  Swiss NSF and the NCCR Nanocience.
\end{acknowledgments}

\bibliography{lsf}

\begin{thebibliography}{9}
\expandafter\ifx\csname natexlab\endcsname\relax\def\natexlab#1{#1}\fi
\expandafter\ifx\csname bibnamefont\endcsname\relax
  \def\bibnamefont#1{#1}\fi
\expandafter\ifx\csname bibfnamefont\endcsname\relax
  \def\bibfnamefont#1{#1}\fi
\expandafter\ifx\csname citenamefont\endcsname\relax
  \def\citenamefont#1{#1}\fi
\expandafter\ifx\csname url\endcsname\relax
  \def\url#1{\texttt{#1}}\fi
\expandafter\ifx\csname urlprefix\endcsname\relax\def\urlprefix{URL }\fi
\providecommand{\bibinfo}[2]{#2}
\providecommand{\eprint}[2][]{\url{#2}}

\bibitem[{fs()}]{fs}
\bibinfo{note}{V. T. Petrashov, I. A. Sosnin, I. Cox, A. Parsons, and C.
  Troadec, Phys. Rev. Lett. {\bf 83}, 3281 (1999); M. Giroud, H. Courtois, K.
  Hasselbach, D. Mailly, and B. Pannetier, Phys. Rev. B {\bf 58}, 11872 (1998);
  V. I. Fal'ko, C. J. Lambert, and A. F. Volkov, Pis'ma Zh. Eksp. Teor. Fiz.
  {\bf 69}, 497 (1999), {JETP} Lett. {\bf 69} 532 (1999); F. J. Jedema, B. J.
  van Wees, B. H. Hoving, A. Filip, and T. M. Klapwijk, Phys. Rev. B {\bf 60},
  16549 (1999); D. Huertas-Hernando, Yu. V. Nazarov, and W. Belzig, Phys. Rev.
  Lett. {\bf 88}, 047003 (2002)}.

\bibitem[{\citenamefont{Gu et~al.}(2002)\citenamefont{Gu, Caballero, Slater,
  Loloee, and Pratt}}]{Gu:prb02}
\bibinfo{author}{\bibfnamefont{J.~Y.} \bibnamefont{Gu}},
  \bibinfo{author}{\bibfnamefont{J.~A.} \bibnamefont{Caballero}},
  \bibinfo{author}{\bibfnamefont{R.~D.} \bibnamefont{Slater}},
  \bibinfo{author}{\bibfnamefont{R.}~\bibnamefont{Loloee}}, \bibnamefont{and}
  \bibinfo{author}{\bibfnamefont{W.~P.} \bibnamefont{Pratt},
  \bibfnamefont{Jr.}}, \bibinfo{journal}{Phys. Rev. B}
  \textbf{\bibinfo{volume}{66}}, \bibinfo{pages}{140507}
  (\bibinfo{year}{2002}).

\bibitem[{\citenamefont{Belzig et~al.}(1999)\citenamefont{Belzig, Wilhelm,
  Bruder, Sch\"{o}n, and Zaikin}}]{Belzig:sm99}
\bibinfo{author}{\bibfnamefont{W.}~\bibnamefont{Belzig}},
  \bibinfo{author}{\bibfnamefont{F.~K.} \bibnamefont{Wilhelm}},
  \bibinfo{author}{\bibfnamefont{C.}~\bibnamefont{Bruder}},
  \bibinfo{author}{\bibfnamefont{G.}~\bibnamefont{Sch\"{o}n}},
  \bibnamefont{and} \bibinfo{author}{\bibfnamefont{A.~D.}
  \bibnamefont{Zaikin}}, \bibinfo{journal}{Superlatt. Microstruc.}
  \textbf{\bibinfo{volume}{25}}, \bibinfo{pages}{1251} (\bibinfo{year}{1999}).

\bibitem[{\citenamefont{Kopnin}(2001)}]{Kopnin:01}
\bibinfo{author}{\bibfnamefont{N.}~\bibnamefont{Kopnin}},
  \emph{\bibinfo{title}{Theory of {N}onequilibrium {S}uperconductivity}}
  (\bibinfo{publisher}{Oxford {S}cience {P}ublications}, \bibinfo{year}{2001}).

\bibitem[{\citenamefont{Rammer and Smith}(1986)}]{Rammer:rmp86}
\bibinfo{author}{\bibfnamefont{J.}~\bibnamefont{Rammer}} \bibnamefont{and}
  \bibinfo{author}{\bibfnamefont{H.}~\bibnamefont{Smith}},
  \bibinfo{journal}{Rev. {M}od. {P}hys.} \textbf{\bibinfo{volume}{58}},
  \bibinfo{pages}{323} (\bibinfo{year}{1986}).

\bibitem[{\citenamefont{Tserkovnyak and Brataas}(2002)}]{Tserkovnyak:prb02}
\bibinfo{author}{\bibfnamefont{Y.}~\bibnamefont{Tserkovnyak}} \bibnamefont{and}
  \bibinfo{author}{\bibfnamefont{A.}~\bibnamefont{Brataas}},
  \bibinfo{journal}{Phys. Rev. B} \textbf{\bibinfo{volume}{65}},
  \bibinfo{pages}{094517} (\bibinfo{year}{2002}).

\bibitem[{Mae()}]{Maekawa}
\bibinfo{note}{T. Yamashita, S. Takahashi, H. Imamura, and S. Maekawa, Phys.
  Rev. B {\bf 65}, 172509 (2002); S. Takahashi and S. Maekawa, Phys. Rev. B
  {\bf 67}, 052409 (2003)}.

\bibitem[{\citenamefont{{S}chmid}(1981)}]{Schmid:nato81}
\bibinfo{author}{\bibfnamefont{A.}~\bibnamefont{{S}chmid}}
  (\bibinfo{year}{1981}), vol.~\bibinfo{volume}{65} of
  \emph{\bibinfo{series}{{NATO} {A}dvanced {S}tudy {I}nstitute {S}eries {B}}},
  pp. \bibinfo{pages}{423--480}.

\bibitem[{spi()}]{spininjection}
\bibinfo{note}{M. Johnson and R. H. Silsbee, Phys. Rev. Lett. {\bf 55}, 1790
  (1985); F. J. Jedema, A. Filip, and B. J. van Wees, Nature (London) {\bf
  410}, 345 (2001)}.

\end{thebibliography}

\end{document}